\documentclass[12pt,onecolumn]{IEEEtran}
\pdfoutput=1
\usepackage[T1]{fontenc}
\usepackage{amsmath}
\usepackage{graphicx}
\usepackage{amssymb}
\usepackage{gensymb}
\usepackage{filecontents,lipsum}
\usepackage{float}
\usepackage[noadjust]{cite}
\usepackage{scrtime}

\begin{document}

\newcommand*{\QEDA}{\hfill\ensuremath{\blacksquare}}
\newcommand*{\QEDB}{\hfill\ensuremath{\square}}
\title{Consensus tracking in multi agent system with nonlinear and non identical dynamics via event driven sliding modes}

\author{Abhinav Sinha and Rajiv Kumar Mishra

\thanks{Rajiv Kumar Mishra is with School of Electronics Engineering, KIIT University, India. \textbf{email:} rajivmishra86@gmail.com}
\thanks{Abhinav Sinha is with School of Mechatronics and Robotics, Indian Institute of Engineering Science and Technology, Shibpur (Howrah). \textbf{email:}abhinavsinha876@gmail.com}
\thanks{This manuscript was originally created on July 9, `2017.}
}
\maketitle

\begin{abstract}
In this work, leader follower consensus objective has been addressed with the synthesis of an event based controller utilizing sliding mode robust control. The schema has been partitioned into two parts viz. finite time consensus problem and event triggered control mechanism. A nonlinear multi agent system with non identical dynamics has been put forward to illustrate the robust capabilities of the proposed control. The first part incorporates matching of states of the followers with those of the leader via consensus tracking algorithm. In the subsequent part, an event triggered rule is devised to save computational power and restrict periodic updating of the controller involved while ensuring desired closed loop performance of the system. Switching of the event based controller is achieved via sliding mode control. Advantage of using switched controller like sliding mode is that it retains its inherent robustness as well as event triggering approach aids in saving energy expenditure. Efficacy of the proposed scheme is confirmed via numerical simulations.
\end{abstract}


\section{Introduction}
\IEEEPARstart{R}{ecently} distributed cooperative control for Multi-Agent Systems (MAS) has attracted researchers from control and allied sciences fraternity from around the globe because of having potential application across various fields like formation control of robots, intelligent transportation systems, attitude alignment of spacecraft, cooperative attack of multiple missiles, unmanned aerial vehicles (UAV) \cite{7192889uav1} for rescue operation, distributed computing and distributed sensor networks, etc. In an interconnection of several autonomous agents, a perturbation in a single agent can affect the entire network. The decentralized control approach has been proved effective to control such systems. A high degree of autonomy can only be achieved if each of the agents can be controlled via robust design techniques, and at the same time, an efficient coordination control is also possible to reach a common goal. Several autonomous agents form a class of dynamic system where they interact with each other over a communication network to reach a common goal. In comparison to a single autonomous agent, MAS provides higher degree of redundancy and improves the operational efficiency. A particularly challenging problem in this field is commonly known as consensus tracking, i.e., consensus with a leader (agent in swarm whose dynamics has to be imitated by others). In this problem, a group of followers tracks the states of a virtual leader in finite time using only local interactions. A group of agents should possess the ability to communicate with their neighbours and are required to accomplish some tasks together as a single entity. If all the agents in the group can be made to come to a common ground, the agents are said to have reached consensus or an agreement. This means the agents have achieved consensus about some state dependent parameter of interest to meet a requirement collectively. A consensus algorithm or protocol is an interaction rule that specifies the information exchange between an agent and all of its neighbors on the network \cite{ref10}. Another challenging problem is referred to as formation control. The objective of the formation control is to stabilize the relative distance or velocity amongst agents to a given value. In this work, the leader has its dynamics independent of the followers and the followers are required to track the leader's state in spite of the presence of uncertainties and disturbances. The communication amongst these agents is achieved via wireless means connected over a shared network with limited resources. In such scenario, continuous updating of the control signal can consume more energy. Hence, an event triggering approach is incorporated which updates the control signal only when some condition (predefined) gets violated. This aids in the efficient utilization of shared resources, less computational power expenditure and increases the life of hardware as well. Event based schema involves switching every time a predefined threshold gets crossed over. This switching is implemented by a sliding mode controller known for its inherent disturbance rejection capabilities. \\

Observations in nature such as a flock of birds, a school of fish, and a swarm of insects have led to the development of algorithms for collective behavior of MAS. The first flocking behavior was simulated by Reynolds \cite{ref22}. Subsequently, there are various early contributions in this field. A brief literature survey goes here.\\

 In \cite{ref111}, the consensus problem of second order MAS has been addressed based on the graph having directed spanning tree. A Laplacian matrix based approach to observing the flow of information amongst agents has been proposed in \cite{ref12} to investigate cooperative control of a group of agents modelled as a network. The solution to the problem of linear leader follower multi agent networks has been proposed in \cite{ref13} based on observer design. The dynamic consensus tracking for directed network topology using proportional- derivative algorithm has been proposed in \cite{ref14}. The nonlinear dynamics of MAS has been approximated using adaptive neural network \cite{ref15} and fuzzy logic \cite{ref16} and consensus has been investigated. The leader follower consensus control has also been proposed where dynamics of a virtual leader is controlled and followers are made to follow desired trajectory \cite{ref17}. The consensus problem of a second order MAS with directed communication topology, as opposed to undirected communication topology has also been addressed in literature using terminal sliding mode control \cite{ref18}. The sliding mode controllers are applied for finite time distributed tracking of second order MAS in \cite{ref20}. Moreover, an observer based event triggered consensus problem has been investigated for position tracking of MAS in \cite{ref21}. The containment control problem for Euler-Lagrange systems has been addressed in \cite{ref24}. The Distributed $H_\infty$ consensus and control has been investigated in \cite{ref23}. In \cite{7835268}, a distributed finite time consensus control for higher order MAS has been dealt with.\\

Event based sampling and control has become popular in energy constrained applications. As opposed to periodic update of controller, control signal is updated only when the measurement error crosses a predefined threshold. The event based broadcasting of state information for multi agent coordination control has been described in \cite{SEYBOTH2013245}. Event triggered leader follower tracking control for MAS has been investigated in \cite{ref25}. An event based controller for rendezvous problem of MAS has been studied in \cite{FAN2013671}.\\

A novel event driven sliding mode control for distributed cooperative control of MAS having non-identical dynamics has been addressed in this work. The paper is organized as follows. Section II introduces preliminaries of graph theory and event triggered sliding mode control. Governing dynamics of the multi agent system has been discussed in section III, followed by the synthesis of proposed controller in section IV. Numerical simulations are given in section V. Section VI provides the concluding remarks.

\section{Preliminaries}
A brief background that is necessary for the foregoing discussion is presented in this section. A familiarity with spectral graph theory and event based sliding mode control is introduced.

\subsection{Spectral Graph Theory applied to Multi Agent Systems}
A directed graph, also known as digraph \cite{Gary} is represented here by $\mathcal{G} = (\mathcal{V,E,A})$. $\mathcal{V}$ is the nonempty set that contains finite number of vertices or nodes \cite{Deo:1974:GTA:1096898}, \cite{mgt} such that $\mathcal{V} = \{1, 2, ..., N\}$. $\mathcal{E}$ denotes edges that are directed and  represented as $\mathcal{E} = \{(i,j)\hspace{1mm} \forall \hspace{1mm} i,j \in \mathcal{V} \hspace{1mm}\&\hspace{1mm} i\neq j\}$. $\mathcal{A}$ is the weighted adjacency matrix such that $\mathcal{A} = a(i,j) \in \mathbb{R}\textsuperscript{N $\times$ N}$.\\
The existence of an edge $(i,j)$ is only possible if and only if the vertex $i$ receives the information supplied by the vertex $j$, i.e., $(i,j) \in \mathcal{E}$. Hence, $i$ and $j$ are referred to as neighbours. Let us consider a set $\mathcal{N}_i$ that contains labels of vertices that are neighbour of the vertex $i$. For the adjacency matrix $\mathcal{A}$, $a(i,j) \in \mathbb{R}\textsuperscript{+} \cup \{0\}$. If $(i,j) \in \mathcal{E} \Rightarrow a(i,j)>0$. If $(i,j) \notin \mathcal{E}$ or $i = j \Rightarrow a(i,j)=0$.\\

The Laplacian matrix $\mathcal{L}$ \cite{chung}, \cite{zhang}, \cite{Gary} lies at the heart of the consensus problem and is given by $\mathcal{L} = \mathcal{D-A}$ where $\mathcal{D}$ is a diagonal matrix, i.e, $\mathcal{D}$ = diag($d_1,d_2,...,d_n$) whose entries are $d_i = \sum_{j=1}^{n} a(i,j)$. A directed path from vertex $j$ to vertex $i$ defines a sequence comprising of edges $(i,i_1), (i_1,i_2), ..., (i_l,j)$ with distinct vertices $i_k \in \mathcal{V}$, $k = 1, 2, 3, ..., l$. $\mathcal{B}$ is also a diagonal matrix with entries $1$ or $0$. If there exists an edge between leader agent and any other agent, the entry is $1$ and $0$ otherwise. Furthermore, it can be inferred that the path between two distinct vertices is not uniquely determined. However, if a distinct node in $\mathcal{V}$ contains directed path to every other distinct node in $\mathcal{V}$, then the directed graph $\mathcal{G}$ is said to have a spanning tree.\\
Physically, each agent in the multi agent system is represented by a vertex or node and the line of communication between any two agents is represented as a directed edge. The relationship between $\mathcal{G}$ and $\mathcal{V}$ establishes the following lemmas.\\\\
\textbf{Lemma 1}: Consider a directed graph $\mathcal{G}$ and its Laplacian matrix $\mathcal{L}$. The set of eigenvalues of $\mathcal{L}$ contains at least one zero eigenvalue. Other nonzero eigenvalues of $\mathcal{L}$ have positive real parts. $\mathcal{L}$ has a simple zero eigenvalue only when $\mathcal{G}$ has a spanning tree. Also, $\mathcal{G}$ is said to be balanced if the following criterion is met.
\begin{equation}
\mathcal{L}\bold{1}_N = \bold{1}_N^T \mathcal{L} = \bold{0}_N
\end{equation}
where $\bold{1}_N$ denotes a column vector of all 1s, i.e, $[1, 1, ..., 1]^T$ and $\bold{0}_N$ denotes a column vector of all 0s, i.e, $[0, 0, ..., 0]^T$. Both $\bold{1}_N$ and $\bold{1}_N$ $\in \mathbb{R}\textsuperscript{N}$. If elements of $\mathcal{L}$ be denoted as $l(i,j)$, then $l(i,j) \in \mathbb{R}\textsuperscript{N $\times$ N}$\\

\textbf{Lemma 2}: The matrix $\mathcal{L} + \mathcal{B}$ has full rank when $\mathcal{G}$ has a spanning tree with leader as the root. This implies non singularity of $\mathcal{L} + \mathcal{B}$ and is used in the synthesis of controller in later stages.

\subsection{Event triggered Sliding Mode Control}
Sliding Mode Control (SMC) \cite{utkin}, \cite{utkin2} is known for its inherent robustness. The switching nature of the control is used to nullify bounded disturbances and matched uncertainties. The switching happens about a surface (hyperplane) in state space known as sliding surface (hyperplane). The control forces the system monotonically towards the sliding surface and this phase is regarded as reaching phase. When the system reaches the sliding surface it remains there for all future time, thereby ensuring that the system dynamics remain independent of bounded disturbances and matched uncertainties. The controller has reaching phase (trajectories in phase plane emanate and move towards the switching surface) and sliding phase (trajectories in the phase plane that reach the switching surface try to remain there).

\subsubsection{Reaching Phase}
Let the hyperplane discussed be given as $\sigma(x)$. In order to drive state trajectories onto this manifold, a proper discontinuous control effort $u(t,x)$ needs to be synthesized that satisfies the following inequality.
\begin{equation}
\sigma^T(x)\dot{\sigma}(x) \leq -\eta \|\sigma(x)\|
\end{equation}
with $\eta$ being positive and is called the reachability constant.\\
$\because$
\begin{equation}
\dot{\sigma}(x) = \frac{\partial \sigma}{\partial x} \dot{x} = \frac{\partial \sigma}{\partial x} f(t,x,u)
\end{equation}
$\therefore$
\begin{equation}
\sigma^T(x) \frac{\partial \sigma}{\partial x} f(t,x,u) \leq -\eta \|\sigma(x)\|
\end{equation}
It is clear that the control $u$ can be synthesized from the above equation.

\subsubsection{Sliding Phase}
The motion of state trajectories confined on the switching manifold is known as \emph{sliding}. A sliding mode is said to exist if the state velocity vectors are directed towards the manifold in its neighbourhood \cite{zak}, \cite{utkin}. Under this circumstance, the manifold is called attractive \cite{zak}, i.e., trajectories starting on it remain there for all future time and trajectories starting outside it tends to it in an asymptotic manner.
\begin{align}
\because \hspace{3mm}\dot{\sigma}(x) = \frac{\partial \sigma}{\partial x} f(t,x,u) \nonumber \\
\text{Hence, in sliding motion} \nonumber \\
\frac{\partial \sigma}{\partial x} f(t,x,u) = 0
\end{align}
Then $u = u_{eq}$ (say) be a solution and is generally referred to as the equivalent control. This $u_{eq}$ is not the actual control applied to the system but can be thought of as a control that must be applied on an average to maintain sliding motion. It is mainly used for analysis of sliding motion \cite{yan2017}. \\

In practice, individual autonomous agents in MAS are often equipped with small digital microcontrollers to reduce the cost. These microcontrollers have limited computing and communication capabilities. According to traditional sampled data control systems theory, samples of a measured output are obtained in periodic fashion with a fixed sampling rate. In addition to this a zero order hold operator is needed to maintain the control input signal constant between successive sample instants. This sampling technique is known as periodic sampling and is generally done along the time axis, also known as \emph{Riemann} sampling \cite{1184824}. An alternative and more efficient way to obtain samples along dependent variable axis (vertical axis), known as \emph{Lebesgue} sampling \cite{1184824}. Under this technique, sampling interval is no longer periodic and samples are obtained only when a \emph{noticeable change}, also referred to as an \emph{event} occurs. As a result, the controller does not need to update itself periodically, but a hold type operator is still needed to maintain the value constant between successive sample instants. The continuous sampling and transmission, along with the occupancy of central processing unit to perform computations when the signal is constant (not changing too frequently) lead to significant waste of available resources. The optimum utilization of communication, computing and energy expenses is a concern  in various applications with increasing number of systems getting networked. One mitigation strategy adopted is event based control wherein control is applied only when the system calls for it depending upon some \emph{event}. Event based sampling is a tradeoff between performance and sampling frequency. More formal presentation and earlier contributions on event based control are presented in \cite{marconi}, \cite{behera}, \cite{lingshi}, \cite{mazo}, \cite{tabuda}, \cite{Aström2008}, \cite{anta}, \cite{chopra}, \cite{lemmon2010}.\\

As a consequence of combining event based strategies with sliding mode control, the robustness of the system has been retained while maintaining lower computational expense. However, the system trajectories tend to move away from the sliding manifold till the control is updated again but remain bounded within a band. Detailed discussion has been carried out in later sections.

\section{System Dynamics}
Let us take into account a multi agent system with a virtual leader and a finite number of followers that are interconnected in a well defined topology. Under this topology, information of the leader's state is not available globally. However,  local information is obtained by communication between follower agents.\\
In such a system, the dynamics of the leader and followers are given by a nonlinear differential equation described below.
\begin{equation}
\dot{x}_0 = f(t,x_0) + u_0 + \varsigma_0
\end{equation}
\begin{equation}
\dot{x}_i = f(t,x_i) + u_i + \varsigma_i \hspace{5mm};\hspace{5mm} i \in \mathbb{N}
\end{equation}
 where $f:\mathbb{R}\textsuperscript{+} \times X \rightarrow \mathbb{R}\textsuperscript{m}$ denotes the uncertain nonlinear dynamics of each agent in the multi agent system. Here the map $f$ is taken to be continuous in $t$. Also $X \subset \mathbb{R}\textsuperscript{m}$ is a domain in which origin is contained.\\
$x_0$ represents the state of the virtual leader and $u_0$ is the control associated with it. Also $u_0$ satisfies $\|u_0\| \leq \Lambda$ for some $\Lambda \in \mathbb{R}\textsuperscript{+}$. This makes quite a practical case when upper limits on hardware constraints are known but the information on real time control effort is not concrete. Quite similarly $x_i$ and $u_i$ are state of $i$\textsuperscript{th} follower and the associated control respectively. $\varsigma_0$ and $\varsigma_i$ are bounded disturbances that may creep in the system.\\
In this study, the dynamics of the leader is independent of that of the followers. Moreover, the dynamics of follower agents considered need not be identical. Quite generally, every agent whether leader or any follower can have a dynamics totally different from the other. In literature, such agents are referred as heterogenous agents. Furthermore, it is assumed throughout the discussion that the functions described by $(6)$ and $(7)$ are locally Lipschitz on some fairly large domain $\mathbb{D_L}$ with Lipschitz contsant $\bar{L}$, i.e.,
\begin{equation}
\|f(x_1) - f(x_2)\| \leq \bar{L} \|x_1 - x_2\|
\end{equation}
$\forall t \in \mathbb{R}\textsuperscript{+} \cup \{0\}$; $x_1$, $x_2 \in X$ and $\bar{L}$ $\in \mathbb{R}\textsuperscript{+}$

\section{Synthesis of the proposed controller}
The prime focus of this work is to ensure accurate trajectory tracking of the leader agent by other agents in finite time with minimum computational expense. The consensus tracking aims to maintain follower's state in consistence with leader's state in finite time by local communication. However, the leader moves independently of followers.\\
Let us define the tracking error for $i^{th}$ agent as
\begin{equation}
e_i(t) = x_i(t) - x_0(t)
\end{equation}
In terms of graph theory, the error candidate modifies to \cite{hindawi}
\begin{equation}
\bar{e_i}(t) =(\mathcal{L+B})  e_i(t) = (\mathcal{L+B})  (x_i(t) - x_0(t))
\end{equation}
It is required that this error candidate vanishes quickly to ensure a common agreement between agents and they attain accurate tracking. Furthermore, formation control is also addressed in this work to demonstrate one possible application. The error candidate for formation control can be suitably modified as
\begin{equation}
\bar{e_i}(t) = (\mathcal{L+B})  (x_i(t) - x_0(t) - \alpha_i)
\end{equation}
where $\alpha_i$ denotes the distance between leader and $i^{th}$ follower.\\

The controller that guarantees finite time consensus is synthesized on archetype of variable structure control techniques with a slight modification. For any sliding mode controller design \cite{zak}, \cite{utkin} is a two step process: the design of a sliding surface where trajectories are required to be confined in finite time, and a control to force the trajectories onto this surface. By the theory of sliding modes, the surface variable for $i^{th}$ agent is given as
\begin{equation}
\sigma_i(t) = \bar{e_i}(t) =(\mathcal{L+B})  e_i(t) = (\mathcal{L+B})  (x_i(t) - x_0(t)) \nonumber
\end{equation}
\begin{equation}
\Rightarrow \dot{\sigma}_i(t) =  \dot{\bar{e_i}}(t) =(\mathcal{L+B})  \dot{e}_i(t) =  (\mathcal{L+B})  (\dot{x}_i(t) - \dot{x}_0(t)) \nonumber
\end{equation}
\begin{equation}
\Rightarrow \dot{\sigma}_i(t) = (\mathcal{L+B}) ( f(t,x_i) + u_i(t) + \varsigma_i - f(t,x_0) - u_0(t) - \varsigma_0) \nonumber
\end{equation}
\begin{equation}
\Rightarrow \dot{\sigma}_i(t) = (\mathcal{L+B}) ( f(x_i(t)) + u_i(t) + \varsigma_i - f(x_0(t)) - u_0(t) - \varsigma_0) \nonumber
\end{equation}
\begin{multline}
\Rightarrow u_i(t) = -(\mathcal{L+B})^{-1}(K |\sigma_i(t)|^\tau sign(\sigma_i(t)) + f(x_i(t)) \\
- f(x_0(t)) - u_0(t) + \varsigma_i -  \varsigma_0)
\end{multline}
where $K |\sigma_i(t)|^\tau sign(\sigma_i(t))$ is the discontinuous forcing function and $K$ is adjustable gain which can be tuned as per design needs. The exponent $\tau \in (0,1)$ and is a design parameter facilitating additional tuning. \\

The sudden increase of interest in the event driven design of circuits and systems is due to its better performance in applications where resources are constrained. In networked control system like MAS connected over shared network consisting of rapid information exchange between nodes, resources such as bandwidth and processor time are always constrained. In such scenario, event based control is expected to yield better results. The control gets updated only when an event (noticeable change) occurs, thereby significantly minimizing computational requirement and power consumption. The event based control is a good candidate if the requirement is to execute different task in time shared manner and also where control is expensive. It is also advantageous in situations when steady state needs to be upper bounded at start regardless of any initial condition and how the state evolves.\\
It has often been described as an alternative to periodic sampling owing to its nature. Next sample instant is dependent on the triggering of an \emph{event}. Hence, our control law given in $(12)$ is modified for $\forall t \in [t^k, t^{k+1}[$ as below.
\begin{multline}
u_i(t) = -(\mathcal{L+B})^{-1}(K |\sigma_i(t^k)|^\tau sign(\sigma_i(t^k)) + f(x_i(t^k)) \\
- f(x_0(t^k)) - u_0(t^k) + \varsigma_i -  \varsigma_0)
\end{multline}
The error introduced due to discretization of the control is given by
\begin{equation}
\bar{\epsilon}(t) =  x(t^k) - x(t)
\end{equation}
such that at $t^k$, $\bar{\epsilon}(t) = 0$. It should be noted that $t_i^k$ is the triggering instant for $i^{th}$ agent. Note that henceforth $\bar{\epsilon}_0(t)$ and $\bar{\epsilon}_i(t)$ shall correspond to the error described by $(14)$ for leader and follower agents respectively.\\
From $(9)$, $e_i(t) = x_i(t) - x_0(t)$, so $e_i(t^k) = x_i(t^k) - x_0(t^k)$.\\
Similarly, let us define
\begin{equation}
\tilde{e_i}(t) =  e_i(t^k) - e_i(t)
\end{equation}
\textbf{Theorem 1}: Consider the system described by $(6), (7)$, error candidates $(9), (10), (14)$, sliding manifold (hyperplane) $\sigma_i$ in the notions of sliding mode and control law of $(13)$.\\
$(i)$ Sliding mode is said to exist in vicinity of the sliding surface if the surface is an essential attractor. In other words, reachability to the surface is ascertained for some reachability constant $\eta>0$.\\
$(ii)$ The event driven sliding mode control law $(13)$ provides stability to the system in the sense of Lyapunov if gain $K$ is selected as
\begin{equation}
\begin{split}
K &> \sup\{\tilde{\digamma} -\bar{L}\|e_i(t^k)\|+\|\mathcal{H}\| \bar{L} \|\bar{\epsilon_i}(t)\|  \\
&-\|\mathcal{H}\| \bar{L} \|\bar{\epsilon_0}(t)\| +\|\mathcal{H}\| \bar{L}\|e_i(t^k)\|\} 
\end{split}
\end{equation}
\textbf{Proof}: $(i)$ Let us consider a Lyapunov candidate $V$ such that
\begin{equation}
V = \frac{1}{2}\sigma_i^T(t) \sigma_i(t)
\end{equation}
Time derivative of this candidate, given in $(15)$ for $t \in [t^k, t^{k+1}[$ along the state trajectories yield
\begin{equation}
\begin{split}
\dot{V}&=\sigma_i^T(t) \dot{\sigma}_i(t)\\
                 &= \sigma_i^T(t)  \dot{\bar{e_i}}(t) = \sigma_i^T(t) (\mathcal{L+B})  \dot{e}_i(t) \\
		&=  \sigma_i^T(t)  (\mathcal{L+B})  (\dot{x}_i(t) - \dot{x}_0(t))  \\
	    & = \sigma_i^T(t)  (\mathcal{L+B}) ( f(x_i(t)) + u_i(t) + \varsigma_i - f(x_0(t)) \\
&- u_0(t) - \varsigma_0)
\end{split}
\end{equation}
From $(13)$, we can simplify the above expression as
\begin{equation}
\begin{split}
\dot{V}&= \sigma_i^T(t) \{  (\mathcal{L+B}) f(x_i(t)) -  K |\sigma_i(t^k)|^\tau sign(\sigma_i(t^k)) \\ &
- f(x_i(t^k)) + f(x_0(t^k)) + u_0(t^k) - \varsigma_i +  \varsigma_0 \\&
+ (\mathcal{L+B})(\varsigma_i - \varsigma_0) - (\mathcal{L+B}) f(x_0(t)) - (\mathcal{L+B}) u_0(t) \}
\end{split}
\end{equation}
Let $(\mathcal{L+B}) = \mathcal{H}$ (say). It should be noted that $u_0(t)$ is bounded and leader is not triggered. Also the disturbances are upper bounded by a finite positive quantity.\\
$\therefore$ we can write $(19)$ as
\begin{equation}
\begin{split}
 \dot{V}&= \sigma_i^T(t)\{\digamma - K |\sigma_i(t^k)|^\tau sign(\sigma_i(t^k)) \\ 
&+ (\mathcal{H}) \{f(x_i(t)) - f(x_0(t)) \} - \{f(x_i(t^k)) - f(x_0(t^k)) \} \}
\end{split}
\end{equation}
where $\digamma = u_0(t^k) - (\mathcal{L+B}) u_0(t) + (\mathcal{L+B})(\varsigma_i - \varsigma_0) - (\varsigma_i - \varsigma_0)$ and $\| \digamma \|_\infty \leq \tilde{\digamma}$ for some $\tilde{\digamma}>0$. $(17)$ can be further simplified as
\begin{equation}
\begin{split}
\dot{V}& = \sigma_i^T(t)\{\digamma - K |\sigma_i(t^k)|^\tau sign(\sigma_i(t^k))+ \mathcal{H} f(x_i(t)) \\
&- \mathcal{H} f(x_i(t^k)) + \mathcal{H} f(x_i(t^k)) - \mathcal{H} f(x_0(t)) + \mathcal{H} f(x_0(t^k)) \\
&- \mathcal{H} f(x_0(t^k))- f(x_i(t^k)) + f(x_0(t^k)) \} \\\\
\dot{V}& \leq \sigma_i^T(t)\{\tilde{\digamma} - K |\sigma_i(t^k)|^\tau sign(\sigma_i(t^k))+ \mathcal{H} \{f(x_i(t)) \\
&- f(x_i(t^k))\} + \mathcal{H} f(x_i(t^k)) - \mathcal{H} \{f(x_0(t)) - f(x_0(t^k))\} \\
&- \mathcal{H} f(x_0(t^k)) - f(x_i(t^k)) + f(x_0(t^k)) \} \\\\
\dot{V}& \leq \sigma_i^T(t)\{\tilde{\digamma} - K |\sigma_i(t^k)|^\tau sign(\sigma_i(t^k)) - \bar{L}\|x_i(t^k) - x_0(t^k)\| \\
&+\|\mathcal{H}\| \bar{L} \|x_i(t) - x_i(t^k)\| -\|\mathcal{H}\| \bar{L} \|x_0(t) - x_0(t^k)\| \\
&+ \|\mathcal{H}\| \bar{L}\|x_i(t^k) - x_0(t^k)\| \} \\\\
\dot{V}& \leq \sigma_i^T(t)\{\tilde{\digamma} - K |\sigma_i(t^k)|^\tau sign(\sigma_i(t^k)) - \bar{L}\|e_i(t^k)\| \\
&+\|\mathcal{H}\| \bar{L} \|\bar{\epsilon_i}(t)\| -\|\mathcal{H}\| \bar{L} \|\bar{\epsilon_0}(t)\| + \|\mathcal{H}\| \bar{L}\|e_i(t^k)\| \}\\\\
\dot{V}& \leq - K \sigma_i^T(t)|\sigma_i(t^k)|^\tau sign(\sigma_i(t^k)) + \|\sigma_i^T(t)\| \tilde{\digamma}  \\
&- \|\sigma_i^T(t)\| \bar{L}\|e_i(t^k)\| +\|\sigma_i^T(t)\|\|\mathcal{H}\| \bar{L} \|\bar{\epsilon_i}(t)\| \\
&-\|\sigma_i^T(t)\|\|\mathcal{H}\| \bar{L} \|\bar{\epsilon_0}(t)\| + \|\sigma_i^T(t)\|\|\mathcal{H}\| \bar{L}\|e_i(t^k)\|
\end{split}
\end{equation}
As long as $\sigma_i(t)>0$ or $\sigma_i(t)<0$, the condition $sign(\sigma_i(t)) = sign(\sigma_i(t^k))$ is strictly met $\forall t \in [t^k, t^{k+1}[$. Hence, when trajectories are just outside the sliding surface,
\begin{equation}
\begin{split}
\dot{V}& \leq - K |\sigma_i(t^k)|^\tau \|\sigma_i^T(t)\| + \|\sigma_i^T(t)\| \tilde{\digamma} - \|\sigma_i^T(t)\| \bar{L}\|e_i(t^k)\| \\
&+\|\sigma_i^T(t)\|\|\mathcal{H}\| \bar{L} \|\bar{\epsilon_i}(t)\| -\|\sigma_i^T(t)\|\|\mathcal{H}\| \bar{L} \|\bar{\epsilon_0}(t)\| \\
&+ \|\sigma_i^T(t)\|\|\mathcal{H}\| \bar{L}\|e_i(t^k)\| \\\\
\dot{V}& \leq \|\sigma_i^T(t)\| \{- K |\sigma_i(t^k)|^\tau  + \tilde{\digamma} -\bar{L}\|e_i(t^k)\| \\
&+\|\mathcal{H}\| \bar{L} \|\bar{\epsilon_i}(t)\| -\|\mathcal{H}\| \bar{L} \|\bar{\epsilon_0}(t)\| +\|\mathcal{H}\| \bar{L}\|e_i(t^k)\| \}
\end{split}
\end{equation}
\begin{equation}
\therefore \hspace{3mm} \dot{V} \leq - \eta \|\sigma_i^T(t)\| =  - \eta \|\sigma_i(t)\|
\end{equation}
where $\eta>0$ and $K> \sup\{\tilde{\digamma} -\bar{L}\|e_i(t^k)\|+\|\mathcal{H}\| \bar{L} \|\bar{\epsilon_i}(t)\| -\|\mathcal{H}\| \bar{L} \|\bar{\epsilon_0}(t)\| +\|\mathcal{H}\| \bar{L}\|e_i(t^k)\|\}$.\\
This implies that the sliding manifold is an attractor and trajectory continuously decrease towards it $\forall t \in [t^k, t^{k+1}[$.
This completes the proof of reachability. \QEDA \\\\
$(ii)$ Now, it requires to be shown that if negative definiteness of time derivative of the Lyapunov candidate $(17)$ be ascertained, stability in the sense of Lyapunov can be guaranteed. At instant $t^k$, the control signal gets updated, thereby nullifying the discretization errors. Hence, $\|\bar{\epsilon_i}(t^k)\| \rightarrow 0$ and $\|\bar{\epsilon_0}(t^k)\| \rightarrow 0$\\
When $\sigma_i(t) = 0$ and $t = t^k$, it follows that \\
\begin{equation}
\begin{split}
\sigma_i(t) = \bar{e_i}(t) =(\mathcal{L+B})  e_i(t) = 0 \\
\Rightarrow (\mathcal{L+B})  (x_i(t) - x_0(t)) = 0 \\
\Rightarrow x_i(t) - x_0(t) = 0 \\
\end{split}
\end{equation}
$(23)$ is a direct consequence of \textbf{Lemma 2}, wherein $(\mathcal{L+B})$ is of full rank and thus, is invertible. Hence,
\begin{align}
x_i(t) - x_0(t) = 0  \nonumber \\
\Rightarrow x_i(t^k) - x_0(t^k) = 0 \nonumber \\
\Rightarrow e_i(t^k) = 0
\end{align}
$\because \|\bar{\epsilon_i}(t^k)\| \rightarrow 0$, $\|\bar{\epsilon_0}(t^k)\| \rightarrow 0$ and from $(25)$, we can conclude that $(22)$ can be written as
\begin{equation}
\dot{V} \leq \|\sigma_i^T(t)\| (- K |\sigma_i(t^k)|^\tau  + \tilde{\digamma})
\end{equation}
From the results of part $(i)$, it is clear that $\dot{V} < 0$, concluding that stability in the sense of Lyapunov can be ensured. This concludes the proof of stability. \QEDA\\

The inter event time is denoted as $T_i^k = t^{k+1} - t^k$. As the sampling is not uniform, $T_i^k \neq constant$ $\forall k \in [0,\infty[$. The triggering instants are said to be admissible if $t_i^{k+1} - t_i^k \geq T_i^k$, which implies that there exists a positive lower bound on inter execution time and \emph{Zeno} phenomenon does not exist. Since the control is updated at $t^k$ instants only, for time instants in $ [t^k, t^{k+1}[$, trajectories deviate from the sliding surface. However, this deviation is assumed to be bounded by a small finite quantity.\\
The triggering rule is of particular interest as sampling is done only when this criterion is fulfilled. The main idea is to trigger when state deviates from the stable equilibrium by a designer fixed threshold, i.e, $\|e_i\|>c_0$ or $f(e_i) = \|e_i\| - c_0$. This $c_0$ affects the performance of the closed loop system and represents the ultimate set in which the state vector $x(t)$ remains confined around the stable equilibrium.\\\\
Here, a novel triggering rule is proposed which not only depends on error but also on square of its derivative. Adding a derivative term introduces anticipatory property. Many a times, the direction in which error changes is of little importance than actual rate of change. Hence, the triggering rule used in this work is defined to be
\begin{equation}
g = \|\gamma_1 e_i + \gamma_2\dot{e}_i^2 \| - (c_0 + c_1 e^{-\beta t})
\end{equation}
such that $\gamma_1 > 0$, $\gamma_2 > 0$, $c_0 \geq 0$, $c_1 \geq 0$, $c_0 + c_1 > 0$ and $\beta \in (0,\lambda_2(\mathcal{L}))$\\
Here $\lambda_2(\mathcal{L})$ is the second eigenvalue if all the eigenvalues of $\mathcal{L}$ are arranged in ascending order. This means $\lambda_1(\mathcal{L}) < \lambda_2(\mathcal{L}) < \lambda_3(\mathcal{L}) < ... < \lambda_n(\mathcal{L})$
\\\\
The second term $(c_0 + c_1 e^{-\beta t})$ ensures a finite positive lower bound on inter event execution time and hence there is no \emph{Zeno} behaviour, i.e., no two consecutive events occur at same time.\\
The following theorem establishes the existence of a finite lower bound for inter event execution time for each agent triggered separately.\\\\
\textbf{Theorem 2}: Consider the system described by $(6), (7)$, the control given in $(13)$ and the discretization error as defined in $(14)$. The sequence of triggering instants $\{t_i^{k}\}_{k=0}^\infty$ for each agent respects the triggering rule given in $(27)$. Consequently, \emph{Zeno} phenomenon is not exhibited and the inter event execution time for each agent $T_i^k$ is bounded below by a finite positive quantity.\\\\
\textbf{Proof}: Between $k^{th}$ and $(k+1)^{th}$ sampling instant in the execution of control, the discretization error is non zero. $T_i^k$ is the time it takes the discretization error to rise from $0$ to some finite value. Thus,
\begin{equation}
 \frac{d}{dt}\|\bar{\epsilon_i}(t)\| \leq \|\frac{d}{dt}\bar{\epsilon_i}(t)\| \leq \|\frac{d}{dt}x_i(t)\| \nonumber
\end{equation}
\begin{equation}
\Rightarrow \|\frac{d}{dt}\bar{\epsilon_i}(t)\| \leq \| f(t,x_i) + u_i + \varsigma_i \| \nonumber
\end{equation}
Substituting the control input $(13)$ in the above inequality, we get
\begin{equation}
\begin{split}
\|\frac{d}{dt}\bar{\epsilon_i}(t)\| &\leq \| f(t,x_i) -(\mathcal{L+B})^{-1}\{K |\sigma_i(t^k)|^\tau sign(\sigma_i(t^k))\\ 
&+ f(x_i(t^k))- f(x_0(t^k)) - u_0(t^k) + \varsigma_i -  \varsigma_0\} + \varsigma_i\| \\\\
&\leq \| f(x_i(t)) -(\mathcal{L+B})^{-1}\{K |\sigma_i(t^k)|^\tau sign(\sigma_i(t^k))\\ 
&+ f(x_i(t^k))- f(x_0(t^k)) - u_0(t^k) + \varsigma_i -  \varsigma_0\} + \varsigma_i\| \\\\
&\leq \bar{L}\|x_i(t)\| + \| \mathcal{H}^{-1} \|K + \|\mathcal{H}^{-1}\|\bar{L} \|x_i(t^k)\| \\
&+\|\mathcal{H}^{-1}\|\bar{L}\|x_0(t^k)\| + \|\mathcal{H}^{-1}\|\|\varsigma_i\| + \|\mathcal{H}^{-1}\|\|\varsigma_0\| \\
&+\|\varsigma_i\| \\\\
&\leq \bar{L}(\|x_i(t^k)\| + \|\bar{\epsilon_i}(t)\|) + \| \mathcal{H}^{-1} \|K \\
&+ \|\mathcal{H}^{-1}\|\bar{L} \|x_i(t^k)\| + \mho \\\\
&\leq \bar{L}\|\bar{\epsilon_i}(t)\| + (1 + \|\mathcal{H}^{-1}\|)\bar{L} \|x_i(t^k)\| + \mho \nonumber
\end{split}
\end{equation}
\begin{equation}
\Rightarrow \|\frac{d}{dt}\bar{\epsilon_i}(t)\| \leq \bar{L}\|\bar{\epsilon_i}(t)\| + \Omega \|x_i(t^k)\| + \mho
\end{equation}
where $\mathcal{H} = \mathcal{L+B}$, $\Omega = (1 + \|\mathcal{H}^{-1}\|)\bar{L}$ and $\mho = \mathcal{H}\{\bar{L}\|x_0(t^k)\| + \|\varsigma_i\| + \|\varsigma_0\|\} + \|\varsigma_i\|$\\
For $t \in [t^k, t^{k+1}[$, the solution to this differential inequality can be understood by using Comparison Lemma \cite{khalil} with initial condition $\|\bar{\epsilon_i}(t)\| = 0$. Comparison Lemma \cite{khalil}, \cite{alexander} is particularly useful when information on bounds on the solution is more important than the solution itself.\\
$\because \Omega$ and $\mho$ are always positive, the solution to the differential inequality of $(17)$ qualifies to be a finite positive value \cite{DBLP}. Thus,
\begin{equation}
\|\bar{\epsilon_i}(t)\| \leq \phi(t) \hspace{1mm};\hspace{3mm} \phi(t) \in \mathbb{R^+}
\end{equation}
This implies that $T_i^k$ is always lower bounded by some positive quantity, and hence this concludes the proof. \QEDA \\
Moreover, the results established in this proof hold locally over some fairly large compact domain. Under varied settings, the design parameters can be tuned selectively to attain desired closed loop performance. From $(17)$, as long as $g<0$, next sample is not taken and system is said to deliver acceptable closed loop performance.\\
Iteratively, the triggering sequence can be described as
\begin{equation}
t_i^{k+1} = \text{inf} \{t_i \in [t_i^k, \infty[ \hspace{2mm}: g\geq 0 \}
\end{equation}

A finite but not necessarily constant delay $\Delta$ might occur during sampling and is unavoidable due to hardware characteristics. In such cases the control is maintained constant $\forall t_i \in [t_i^k + \Delta, t_i^{k+1} + \Delta[$. It has been assumed that $\Delta$ is negligible and has been neglected innocuously. Hence for our case, control is constant in the interval $[t_i^k, t_i^{k+1}[$.

\section{Numerical Simulations}
A typical communication topology that is used for information exchange among agents is shown in figure $(1)$. The matrices associated with the topology under consideration are given below.
\begin{figure}[H]
\centering
\includegraphics{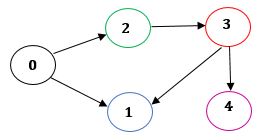}
\caption{Topology of MAS- a digraph with leader indexed with 0 and followers indexed with 1, 2, 3 and 4}
\end{figure}
\begin{equation}
\mathcal{A}=
\begin{bmatrix}
0 & 0 & 1 & 0\\
0 & 0 & 0 & 0\\
0 & 1 & 0 & 0\\
0 & 0 & 1 & 0
\end{bmatrix}
\end{equation}
\begin{equation}
\mathcal{B}=
\begin{bmatrix}
1 & 0 & 0 & 0\\
0 & 1 & 0 & 0\\
0 & 0 & 0 & 0\\
0 & 0 & 0 & 0
\end{bmatrix}
\end{equation}
\begin{equation}
\mathcal{D}=
\begin{bmatrix}
1 & 0 & 0 & 0\\
0 & 0 & 0 & 0\\
0 & 0 & 1 & 0\\
0 & 0 & 0 & 1
\end{bmatrix}
\end{equation}
\begin{equation}
\mathcal{L}=
\begin{bmatrix}
1 & 0 & -1 & 0\\
0 & 0 &  0 & 0\\
0 & -1 & 1 & 0\\
0 & 0 & -1 & 1
\end{bmatrix}
\end{equation}
\begin{equation}
\mathcal{L+B}=
\begin{bmatrix}
2 & 0 & -1 & 0\\
0 & 1 & 0 & 0\\
0 & -1 & 1 & 0\\
0 & 0 & -1 & 1
\end{bmatrix}
\end{equation}
This numerical simulation has been carried out in Mathworks MATLAB\textsuperscript{TM}. Separate cases have been dealt for system operating under influence of an unknown but bounded disturbance, and system operation free from any perturbation. An application to formation control has also been shown to further aid the proposition. The topology shown for demonstration bears Laplacian with real eigenvalues. It is worthy to note that the same discussion applies and can be extended to Laplacian with complex eigenvalues too.\\

For simulation purposes, dynamics of the leader and follower agents considered here are described as
\begin{equation}
\dot{x}_0 = u_0 cos(t) + 0.2 sin(x_0) + \varsigma_0
\end{equation}
where $u_0$ is the control input of the leader and is taken to be
\begin{equation}
u_0 = \frac{2 cos(0.1 \pi t)}{1 + e^{-t}}
\end{equation}
Note that upper bounds on $u_0$ is finite and can be easily calculated.
\begin{align}
\dot{x}_1 = 0.1 sin(x_1)^{1/3} + cos ^ 2 (2 \pi t) + e^{-t}  + u_1(t) + \varsigma_1\\
\dot{x}_2 = 0.1 sin(x_2) + cos  (2 \pi t)   + u_2(t) + \varsigma_2\\
\dot{x}_3 = - x_3 cos (t) - sin (x_3) - cos (x_3)  + u_3(t) + \varsigma_3\\
\dot{x}_4 = sin(x_4) + cos (e^{-x_4 t})  + u_4 (t)+ \varsigma_4
\end{align}
These dynamics are locally Lipschitz over some failrly large domain $\mathbb{D_L}$.
Typical parameter values are tabulated below. These values are valid for all the cases described in this section.
\begin{center}
\begin{tabular}{ |l|l|l|l| }
  \hline 
  \multicolumn{4}{|c|}{\textbf{Numerical values of parameters}} \\
  \hline
  $c_0$ & $10^{-4}$ & $x_1(0)$ & $10$\\
  $c_1$ & $0.2499$ &$x_2(0)$ & $-7$\\
 $\gamma_1$ & $0.8$ &$x_3(0)$ & $4$\\
 $\gamma_2$ & $0.8$ &$x_4(0)$ & $-9$\\
$K$ & 15 &$x_0(0)$ & $0$\\
$\beta$ & 1& $\varsigma_{max}$ matched & (i) $0.3$\\
$\tau$ & 0.5& $\varsigma_{max}$ mismatched & (ii) $9$ \\
  \hline
\end{tabular}
\end{center}

\subsection{System without any external disturbance}
When the system is free from any perturbation, the consensus tracking is achieved in finite time. It is also clear from figure $(2)$ that the convergence to the leader's trajectory is very fast.
\begin{figure}[H]
\centering
\includegraphics[scale=0.4]{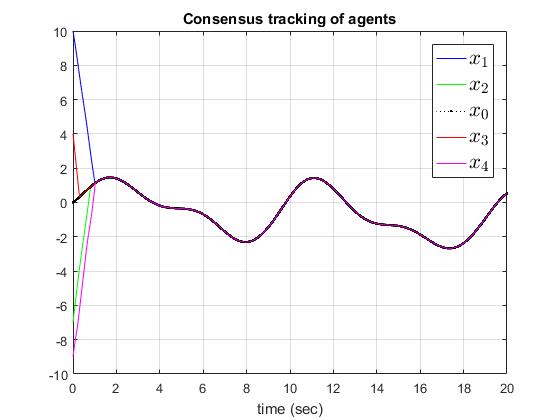}
\caption{Leader- follower consensus in finite time}
\includegraphics[scale=0.4]{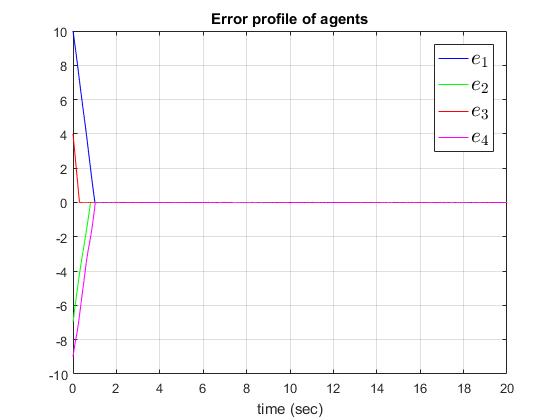}
\caption{Error profile of follower agents}
\includegraphics[scale=0.4]{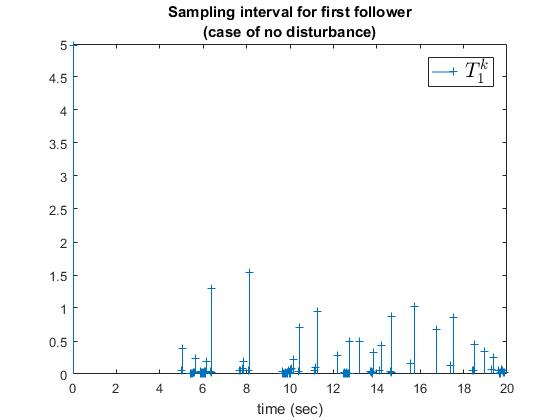}
\caption{Sampling interval - consensus tracking by follower agent 1}
\end{figure}
Figure $(3)$ is the error profile of follower agents. Since the convergence is quite fast, the error variables die out rapidly. Having used a novel triggering rule, the controller updates are minimal. Control effort is required only when necessary. Figures $(4- 7)$  show plots of sampling interval for follower agents. No Zeno behaviour is exhibited. Figure $(8)$  shows the event based sampling instants of each follower agent during consensus. When the sliding mode gets enforced, an increase in sampling is observed.
\begin{figure}[H]
\centering
\includegraphics[scale=0.4]{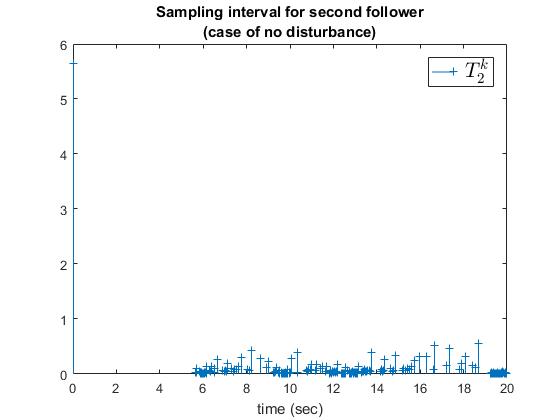}
\caption{Sampling interval - consensus tracking by follower agent 2}
\includegraphics[scale=0.4]{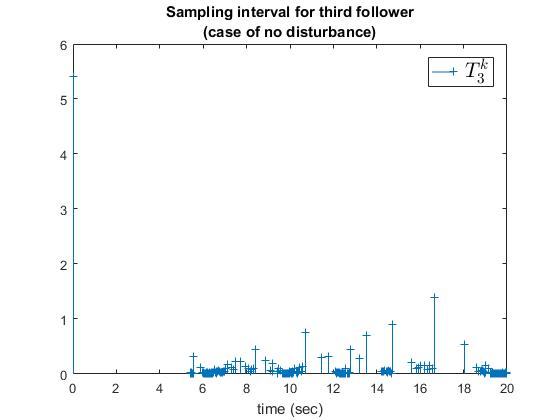}
\caption{Sampling interval - consensus tracking by follower agent 3}
\includegraphics[scale=0.4]{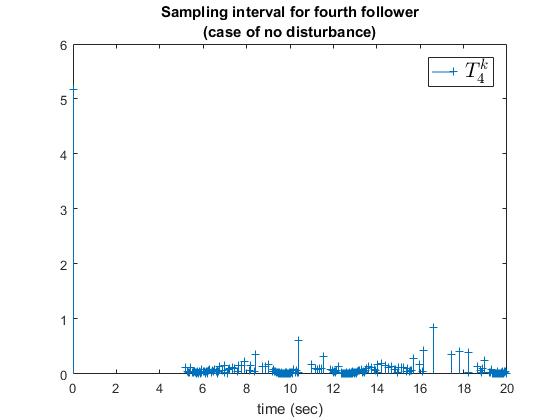}
\caption{Sampling interval - consensus tracking by follower agent 4}
\end{figure}
\begin{figure}
\centering
\includegraphics[scale=0.4]{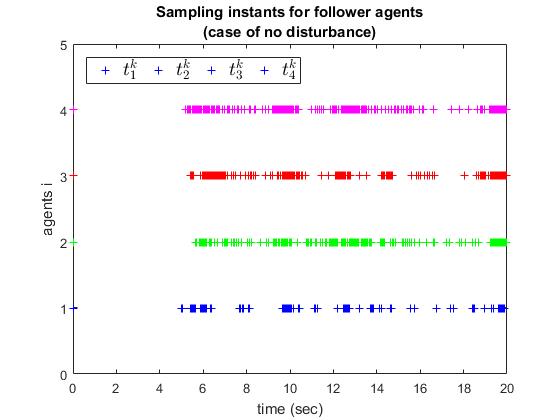}
\caption{Lebesgue sampling instants of each follower agent during consensus}
\end{figure}
Thus, from the plots shown above, it is very clear that consensus tracking has been achieved in finite time. As soon as finite time consensus tracking gets enforced, the error variable reduces to zero. It should be noted that the initial conditions are taken as large values or perturbations far from the stable equilibrium. A similar result follows for initial points very close to the origin.

\subsection{System operating in the presence of external disturbance}
External disturbances may be matched (that are those disturbances which stay within the range space of the input function) or mismatched (that are those disturbances which lie outside the range space of the input function). Sliding mode is known to reject any disturbance that is matched. Here, we show two examples of matched and mismatched perturbations acting on the system but robustness of the closed loop dynamics is not compromised.\\
Let a disturbance acting on the system be of the form
\begin{equation}
\varsigma = \varsigma_{max} \hspace{1mm} sin(\pi^2 t^2) = 0.3\hspace{1mm} sin(\pi^2 t^2)
\end{equation}
tries to corrupt the system dynamics. This disturbance is upper bounded by $0.3$ and is matched. Figure $(9)$ depicts that this disturbance is rejected completely without any ill effect on the closed loop dynamics.
\begin{figure}[H]
\centering
\includegraphics[scale=0.4]{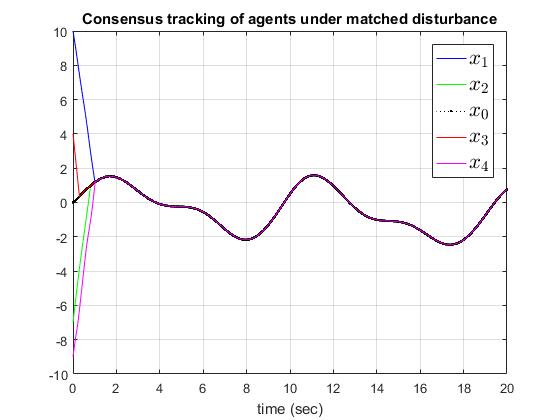}
\caption{Consensus tracking in presence of matched disturbance}
\end{figure}
Consider a mismatched disturbance ($\varsigma_{max} = 9$) affects the system. It can be inferred from figure $(10)$ that although the trajectory gets corrupted, consensus tracking is still achievable in finite time that is not very different from the case when system acts under no disturbance. The initial points are taken as same as that in the previous case. Once again, a good performance has been maintained.
\begin{figure}[H]
\centering
\includegraphics[scale=0.4]{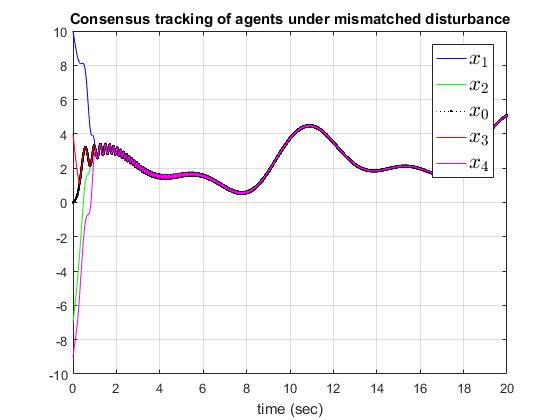}
\caption{Consensus tracking in presence of mismatched disturbance}
\end{figure}

\subsection{Application to formation of agents}
Figure $(11)$ shows a practical application of formation control of MAS. It is evident from the plot that formation is also achieved in finite time and therefore the control is robust.
 \begin{figure}[H]
\centering
\includegraphics[scale=0.4]{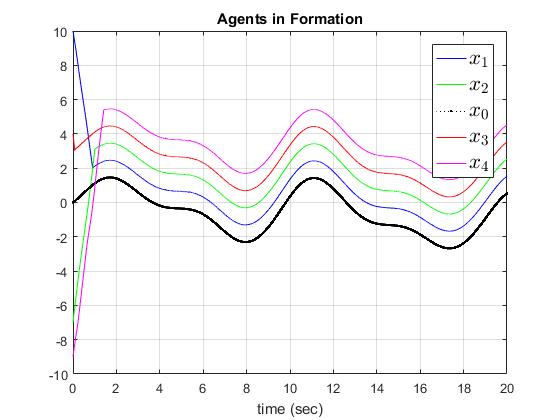}
\caption{Formation of agents in MAS}
\end{figure}
The plot of sampling interval under formation control (figure $(12)$) shows that less computation is required with proposed triggering.
\begin{figure}[H]
\centering
\includegraphics[scale=0.4]{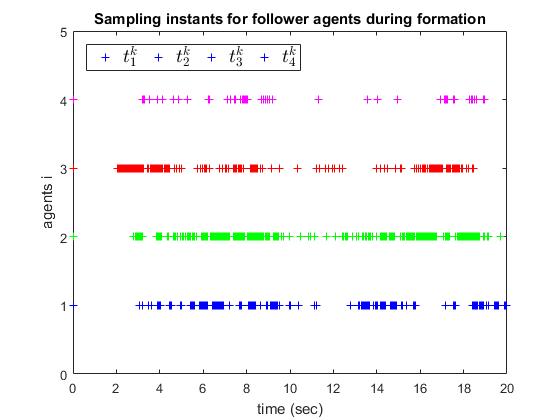}
\caption{Sampling interval - agents in formation}
\end{figure}
Effects of disturbance have also been investigated for formation control. Figure $(13)$ and figure $(15)$ show formation under the influence of matched and mismatched disturbances.
 \begin{figure}[H]
\centering
\includegraphics[scale=0.4]{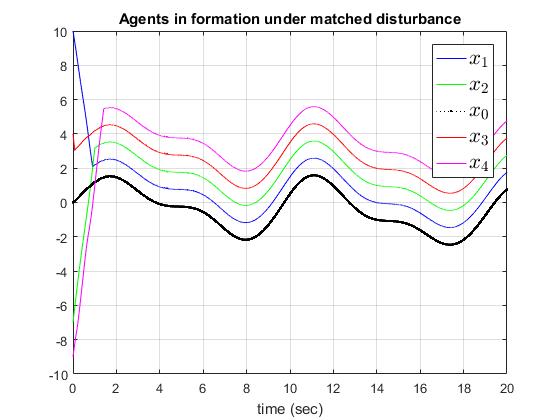}
\caption{Formation of agents in MAS under matched disturbance}
\includegraphics[scale=0.4]{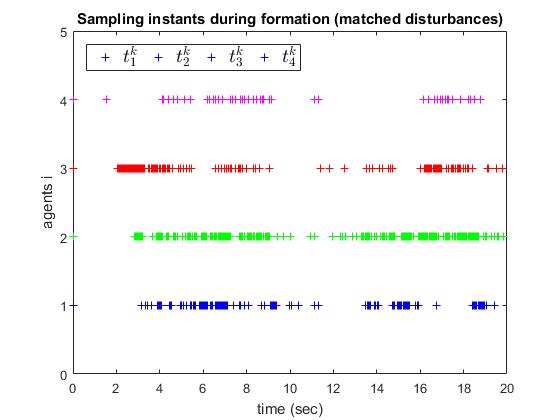}
\caption{Sampling interval - agents in formation under influence of matched disturbance}
\end{figure}
Figure $(14)$ and figure $(16)$ show the sampling instants of each follower agent when the system is influenced by a matched disturbance and a mismatched disturbance respectively of the nature described by $(42)$.
 \begin{figure}[H]
\centering
\includegraphics[scale=0.55]{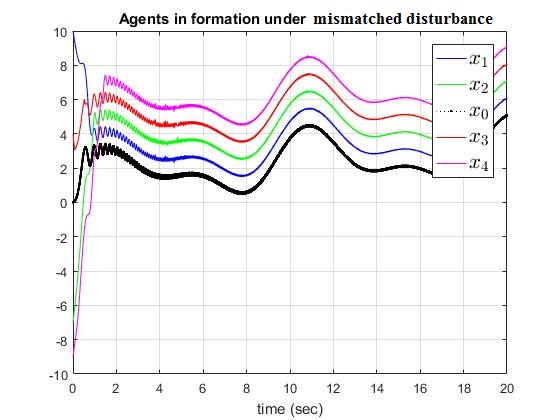}
\caption{Formation of agents in MAS under mismatched disturbance}
\end{figure}
\begin{figure}[H]
\centering
\includegraphics[scale=0.4]{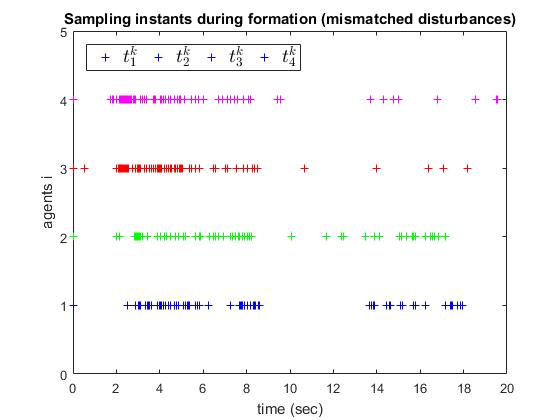}
\caption{Sampling interval - agents in formation under influence of mismatched disturbance}
\end{figure}

\section{Concluding remarks}
The distributed cooperative control problem of leader follower Multi Agent System has been addressed in this work. The dynamics of this MAS is highly nonlinear non identical and hence a sturdy controller is required to drive the agents to a consensus in finite time. The dynamics of this MAS has been treated analytically through basics of directed graphs. The proposed controller has been developed on the notion of sliding mode based on event triggering. This helped achieve low computational power and control by exception. Lyapunov candidate has been chosen such that its negative definiteness proves the stability of the proposed control. If the number of agents on the network is large, then centralized control takes significant computational time and perturbation in an autonomous agent can affect whole network. The added advantage of using sliding mode control is that the robustness of classical sliding mode techniques has been retained and incorporating event triggering strategy helped save computational expenses and power consumption. The inter event execution time is lower bounded by a finite positive value that ensures desired closed loop performance, admissible triggering instants and excludes Zeno phenomenon. Finally, through simulation results, the efficacy of the controller has been demonstrated.

\bibliographystyle{ieeetran}
\bibliography{references}

\begin{thebibliography}{10}
\providecommand{\url}[1]{#1}
\csname url@samestyle\endcsname
\providecommand{\newblock}{\relax}
\providecommand{\bibinfo}[2]{#2}
\providecommand{\BIBentrySTDinterwordspacing}{\spaceskip=0pt\relax}
\providecommand{\BIBentryALTinterwordstretchfactor}{4}
\providecommand{\BIBentryALTinterwordspacing}{\spaceskip=\fontdimen2\font plus
\BIBentryALTinterwordstretchfactor\fontdimen3\font minus
  \fontdimen4\font\relax}
\providecommand{\BIBforeignlanguage}[2]{{%
\expandafter\ifx\csname l@#1\endcsname\relax
\typeout{** WARNING: IEEEtran.bst: No hyphenation pattern has been}%
\typeout{** loaded for the language `#1'. Using the pattern for}%
\typeout{** the default language instead.}%
\else
\language=\csname l@#1\endcsname
\fi
#2}}
\providecommand{\BIBdecl}{\relax}
\BIBdecl

\bibitem{7192889uav1}
{Abhinav Sinha and Rajiv Kumar Mishra}, ``{Robust altitude tracking of a
  miniature helicopter UAV based on sliding mode},'' in \emph{2015
  International Conference on Innovations in Information, Embedded and
  Communication Systems (ICIIECS)}, March 2015, pp. 1--6.

\bibitem{ref10}
{{ S. R. Olfati, J. A. Fax and R. M. Murray}}, ``Consensus and cooperation in
  networked multi-agent systems,'' in \emph{Proc. {IEEE}}, no.~95.

\bibitem{ref22}
C.~W. Reynolds, ``Flocks, herds and schools: A distributed behavioral model,''
  pp. 25--34, 1987.

\bibitem{ref111}
{ W. Ren and E. Atkins}, ``{Distributed multi-vehicle coordinated control via
  local information exchange},'' \emph{International Journal of Robust and
  Nonlinear Control}, no.~17, pp. 1002--1033, 2007.

\bibitem{ref12}
{{ M. Ji and M. Egerstedt}}, ``A graph-theoretic characterization of
  controllability for multi-agent systems,'' in \emph{Proc. {American Control
  Conference}}, 2007, pp. 4588--4593.

\bibitem{ref13}
{ Y. Hong, G. Chen and L. Bushnell}, ``Distributed observers design for leader
  following control of multi-agent networks,'' \emph{Automatica}, no.~44, pp.
  846--850, 2008.

\bibitem{ref14}
{C. Douligeris and G. Develekos}, ``Consensus tracking under direct interaction
  topologies: Algorithms and experiments,'' \emph{IEEE Transactions on Control
  Systems Technology}, no.~18, pp. 230--237, 2010.

\bibitem{ref15}
{Y. J. Liu, S. C. Tong, D. Wang, T. S. Li and C. L. P. Chen}, ``Adaptive neural
  output feedback controller design with reduced-order observer for a class of
  uncertain nonlinear siso systems,'' \emph{IEEE Transactions on Neural
  Networks}, no.~22, pp. 1328--1334, 2011.

\bibitem{ref16}
{Y. J. Liu, S. C. Tong and C. L. P. Chen}, ``Adaptive fuzzy control via
  observer design for uncertain nonlinear systems with un modeled dynamics,''
  \emph{IEEE Transactions on Fuzzy Systems}, no.~21, pp. 275--288, 2013.

\bibitem{ref17}
{X. Lu, R. Lu, S. Chen and J. Lu}, ``Finite-time distributed tracking control
  for multi-agent systems with a virtual leader,'' \emph{IEEE Transactions on
  Circuits and Systems I: Regular Papers}, no.~60, pp. 352--362, 2013.

\bibitem{ref18}
{A. M. Zou, K. D. Kumar and Z. G. Hou}, ``Distributed consensus control for
  multi agent systems using terminal sliding mode and chebyshev neural
  networks,'' \emph{International Journal of Robust and Nonlinear Control},
  no.~23, pp. 334--357, 2013.

\bibitem{ref20}
{C. E. Ren and C. L. P. Chen}, ``Sliding mode leader-following consensus
  controllers for second-order non-linear multi-agent systems,'' \emph{IET
  Control Theory and Application}, no.~9, pp. 1544 -- 1552, 2015.

\bibitem{ref21}
{H. Jiangping, J. Geng and H. Zhu}, ``An observer-based consensus tracking
  control and application to event-triggered tracking,'' \emph{Communications
  in Nonlinear Science and Numerical Simulation}, no.~20, pp. 559--570, 2015.

\bibitem{ref24}
{J. Mei, W. Ren and G. Ma}, ``Distributed containment control for lagrangian
  networks with parametric uncertainties under a directed graph,''
  \emph{Automatica}, no.~48, pp. 653--659, 2012.

\bibitem{ref23}
{P. Lin and Y. M. Jia}, ``{Distributed robust H$\infty$ consensus control in
  directed networks of agents with time-delay},'' \emph{Systems and Control
  Letters}, no.~57, pp. 643--653, 2008.

\bibitem{7835268}
H.~Du, G.~Wen, G.~Chen, J.~Cao, and F.~E. Alsaadi, ``A distributed finite-time
  consensus algorithm for higher-order leaderless and leader-following
  multiagent systems,'' \emph{IEEE Transactions on Systems, Man, and
  Cybernetics: Systems}, vol.~47, no.~7, pp. 1625--1634, July 2017.

\bibitem{SEYBOTH2013245}
\BIBentryALTinterwordspacing
G.~S. Seyboth, D.~V. Dimarogonas, and K.~H. Johansson, ``Event-based
  broadcasting for multi-agent average consensus,'' \emph{Automatica}, vol.~49,
  no.~1, pp. 245 -- 252, 2013. [Online]. Available:
  \url{http://www.sciencedirect.com/science/article/pii/S0005109812004852}
\BIBentrySTDinterwordspacing

\bibitem{ref25}
{Y. Cheng and V. Ugrinovskii}, ``Event-triggered leader-following tracking
  control for multi variable multi-agent systems,'' \emph{Automatica}, no.~70,
  pp. 204--210, 2016.

\bibitem{FAN2013671}
\BIBentryALTinterwordspacing
Y.~Fan, G.~Feng, Y.~Wang, and C.~Song, ``Distributed event-triggered control of
  multi-agent systems with combinational measurements,'' \emph{Automatica},
  vol.~49, no.~2, pp. 671 -- 675, 2013. [Online]. Available:
  \url{http://www.sciencedirect.com/science/article/pii/S0005109812005468}
\BIBentrySTDinterwordspacing

\bibitem{Gary}
{Gary Chartrand, Linda Lesniak, Ping Zhang}, \emph{Graphs \& Digraphs, Sixth
  Edition}, ser. Textbooks in Mathematics.\hskip 1em plus 0.5em minus
  0.4em\relax CRC Press- Taylor and Francis Group, 2015.

\bibitem{Deo:1974:GTA:1096898}
N.~Deo, \emph{Graph Theory with Applications to Engineering and Computer
  Science (Prentice Hall Series in Automatic Computation)}.\hskip 1em plus
  0.5em minus 0.4em\relax Upper Saddle River, NJ, USA: Prentice-Hall, Inc.,
  1974.

\bibitem{mgt}
\BIBentryALTinterwordspacing
{Bollobas Bella}, \emph{Modern Graph Theory}.\hskip 1em plus 0.5em minus
  0.4em\relax Springer- Verlag, 1998. [Online]. Available:
  \url{http://www.springer.com/in/book/9780387984889}
\BIBentrySTDinterwordspacing

\bibitem{chung}
\BIBentryALTinterwordspacing
{Fan R. K. Chung}, \emph{Spectral Graph Theory}, ser. CBMS Regional Conference
  Series in Mathematics.\hskip 1em plus 0.5em minus 0.4em\relax AMS and CBMS,
  1997, vol.~92. [Online]. Available: \url{http://bookstore.ams.org/cbms-92}
\BIBentrySTDinterwordspacing

\bibitem{zhang}
\BIBentryALTinterwordspacing
{Jonathan L. Gross, Jay Yellen, Ping Zhang}, \emph{Handbook of Graph Theory,
  Second Edition}, ser. Discrete Mathematics and Its Applications.\hskip 1em
  plus 0.5em minus 0.4em\relax CRC Press- Taylor and Francis Group, 2013.
  [Online]. Available: \url{http://dx.doi.org/10.1201/b16132}
\BIBentrySTDinterwordspacing

\bibitem{utkin}
{K. David Young, Vadim I. Utkin and Umit Ozguner}, ``A control engineer's guide
  to sliding mode control,'' \emph{IEEE transactions on Control Systems
  Technology}, vol.~7, no.~3, pp. 328--342, May 1999.

\bibitem{utkin2}
{Vadim I. Utkin}, \emph{Sliding Modes in Control and Optimization}.\hskip 1em
  plus 0.5em minus 0.4em\relax Springer, 1992.

\bibitem{zak}
S.~H. $\dot{Z}$ak, \emph{Systems and Control}.\hskip 1em plus 0.5em minus
  0.4em\relax 198 Madison Avenue, New York, New York, 10016: Oxford University
  Press, 2003.

\bibitem{yan2017}
{Yan, Xing-Gang, Spurgeon, Sarah K., Edwards, Christopher}, \emph{Variable
  Structure Control of Complex Systems}.\hskip 1em plus 0.5em minus 0.4em\relax
  Springer, 2017.

\bibitem{1184824}
K.~J. Astrom and B.~M. Bernhardsson, ``Comparison of riemann and lebesgue
  sampling for first order stochastic systems,'' in \emph{Proceedings of the
  41st IEEE Conference on Decision and Control, 2002.}, vol.~2, Dec 2002, pp.
  2011--2016 vol.2.

\bibitem{marconi}
A.~Astholfi and L.~Marconi, \emph{Analysis and design of nonlinear control
  systems (in honour of Alberto Isidori)}.\hskip 1em plus 0.5em minus
  0.4em\relax Springer- Verlag, 2007.

\bibitem{behera}
A.~K. Behera and B.~Bandyopadhyay, ``Event based robust stabilization of linear
  systems,'' in \emph{IECON 2014 - 40th Annual Conference of the IEEE
  Industrial Electronics Society}, Oct 2014, pp. 133--138.

\bibitem{lingshi}
{Dawei Shi, Ling Shi and Tongwen Chen}, \emph{Event based state estimation- A
  Stochastic Perspective}.\hskip 1em plus 0.5em minus 0.4em\relax Springer,
  2016.

\bibitem{mazo}
M.~Mazo and P.~Tabuada, ``Decentralized event-triggered control over wireless
  sensor/actuator networks,'' \emph{IEEE Transactions on Automatic Control},
  vol.~56, no.~10, pp. 2456--2461, Oct 2011.

\bibitem{tabuda}
P.~Tabuada, ``Event-triggered real-time scheduling of stabilizing control
  tasks,'' \emph{IEEE Transactions on Automatic Control}, vol.~52, no.~9, pp.
  1680--1685, Sept 2007.

\bibitem{Aström2008}
K.~J. Astr{\"o}m, \emph{Event Based Control}.\hskip 1em plus 0.5em minus
  0.4em\relax Berlin, Heidelberg: Springer Berlin Heidelberg, 2008, pp.
  127--147.

\bibitem{anta}
A.~Anta and P.~Tabuada, ``To sample or not to sample: Self-triggered control
  for nonlinear systems,'' \emph{IEEE Transactions on Automatic Control},
  vol.~55, no.~9, pp. 2030--2042, Sept 2010.

\bibitem{chopra}
P.~Tallapragada and N.~Chopra, ``On event triggered tracking for nonlinear
  systems,'' \emph{IEEE Transactions on Automatic Control}, vol.~58, no.~9, pp.
  2343--2348, Sept 2013.

\bibitem{lemmon2010}
M.~Lemmon, \emph{Event-Triggered Feedback in Control, Estimation, and
  Optimization}.\hskip 1em plus 0.5em minus 0.4em\relax London: Springer
  London, 2010, pp. 293--358.

\bibitem{hindawi}
\BIBentryALTinterwordspacing
{Nan Liu, Rui Ling, Qin Huang, and Zheren Zhu}, ``Second-order super-twisting
  sliding mode control for finite-time leader-follower consensus with uncertain
  nonlinear multiagent systems,'' \emph{Mathematical Problems in Engineering},
  vol. 2015, p.~8, 2015. [Online]. Available:
  \url{http://dx.doi.org/10.1155/2015/292437}
\BIBentrySTDinterwordspacing

\bibitem{khalil}
{Hasan K. Khalil}, \emph{Nonlinear Systems, Third Edition}.\hskip 1em plus
  0.5em minus 0.4em\relax Upper Saddle River, NJ, USA: Prentice-Hall, Inc.,
  2002.

\bibitem{alexander}
{Alexander G. Ramm, Nguyen S. Hoang}, \emph{Dynamical Systems Method and
  Applications: Theoretical Developments and Numerical Examples}.\hskip 1em
  plus 0.5em minus 0.4em\relax Wiley, 2011.

\bibitem{DBLP}
\BIBentryALTinterwordspacing
A.~Girard, ``Dynamic event generators for event-triggered control systems,''
  \emph{CoRR}, vol. abs/1301.2182, 2013. [Online]. Available:
  \url{http://arxiv.org/abs/1301.2182}
\BIBentrySTDinterwordspacing

\end{thebibliography}

\end{document}